\documentclass{kapproc}

\usepackage{epsfig}

\begin{document}

\articletitle[A First EGRET-UNID-related Agenda for the next-generation IACTs]{A First EGRET-UNID-related Agenda for the next-generation Cherenkov Telescopes}

\author{Dirk Petry}

\affil{Dept. of Physics and Astronomy\\ 
 Iowa State University\\
 Ames, IA 50011}
\email{petry@iastate.edu}


\begin{abstract}
The next generation of Imaging Atmospheric Cherenkov Telescopes (IACTs) will open the regime
between $\approx$ 30 GeV and 200 GeV to ground-based gamma observations with 
unprecedented point source sensitivity and source location accuracy. I examine the prospects 
of observing
the unidentified objects (UNIDs) of the Third EGRET Catalog using the IACT observatories
currently under construction by the CANGAROO, HESS, MAGIC and VERITAS collaborations.
Assuming a modest spectral steepening similar to that observed in the inverse Compton
component of the Crab Nebula spectrum and taking into account the sensitivity of the
instruments and its zenith angle dependence, a detailed list of 78 observable objects 
is derived which is then further constrained to 38 prime candidates. The characteristics 
of this agenda are discussed.
\end{abstract}

\section{Introduction}
The EGRET experiment (1991-1999) has given us the first detailed view of the entire high-energy gamma-ray sky.
Its successful history is described by other authors (e.g. D.J. Thompson, these proceedings).
In parallel with EGRET's observations, observers also began to explore the gamma-ray sky at even
higher energies. The first successful observation of a gamma-ray source above 500 GeV (the Crab Nebula,
Weekes et al. \cite{weekes}) 
using a Cherenkov telescope was made only few years before EGRET's launch and one year 
after it followed the detection of the second source, Mkn 421 (Punch et al. \cite{punch}). Both sources were also 
detected by EGRET and it has become common practice to show the EGRET flux measurement as a reference in plots of 
Cherenkov telescope flux measurements. 

\begin{figure}[ht]
\centerline{\epsfig{file=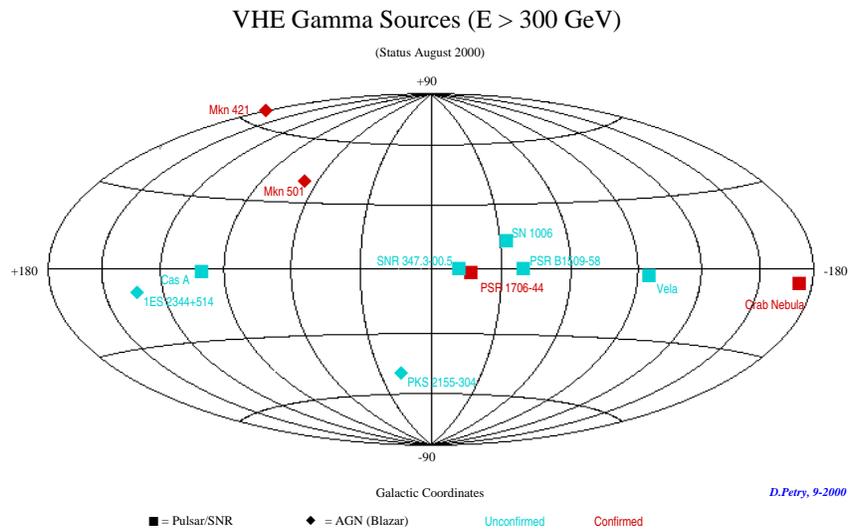, width=\textwidth}}
\caption{ \label{vhemap} A sky map of the presently known VHE gamma-ray sources.  }
\end{figure}

However, it soon became clear that an EGRET detection did not imply that
the source would also be detected at very high energies.
The class of EGRET Blazars has been thoroughly
observed by Cherenkov telescopes, but none has been detected except Mkn 421 although extra\-polations of the EGRET spectra
are in many cases comfortably above the point source sensitivity of the present instruments.
The second extra-galactic source
ever to be detected at very high energies, Mkn 501, was initially not an EGRET source. Only by raising
the energy threshold and looking for flares, Kataoka et al. (\cite{kataoka}) were able to report a marginal
detection. Today, the situation remains essentially the same. About 11 very high energy (VHE) gamma-ray
sources have been detected: 4 Blazars (Mkn 421, Mkn 501, 1ES2344+514, PKS2155-304) and 7 SNRs/Plerions (Crab Nebula,
PSR1706-44, SN1006, Cas A, Vela, PSR B1509-58, SNR 347.3-00.5) (see figure \ref{vhemap}). About half of these are EGRET sources. 

We are therefore looking at three classes of very high energy sources: (a) steeply cutting-off 
sources which are detected by EGRET but become unobservable above a few 100 GeV, (b) flat spectrum sources like Mkn 501
which only become observable at higher energies, and (c) intermediate cases like Mkn 421 or the Crab Nebula.
Most known sources must belong to type (a) given the way observation technology has developed.
Whether this is really a selection effect is unclear. It seems that
the universe becomes abruptly very much darker above a few 100~GeV. 
The number of detectable sources decreases rapidly with rising energy threshold
even though the point source sensitivity of present Cherenkov telescopes is sufficiently many orders of 
magnitude better than that of EGRET to allow for a spectral index of 2.1 in differential photon flux. 
In other words, the cut-off of most high-energy source spectra seems to take place in the
range 1 - 200 GeV. The first decade of energy in this range has been covered by EGRET and
57 sources have been detected (Lamb \& Macomb \cite{onegev}). The remaining 10-200 GeV have never been explored
until today and it is this ``gap'' that forms the major incentive behind the projects for future,
more sensitive Cherenkov telescopes.

According to the Third EGRET Catalog (Hartman et al. \cite{egret3rdcat}), 197 out of the 271
objects in the catalog could so far not be identified with an optical, radio or X-ray counterpart 
with certainty. These sources are the Unidentified EGRET objects (UNIDs), 38 of which have already
a tentative identification. The main reason for the non-identification of so many sources is
the fact that their position is only poorly known because of EGRET's wide point spread function
and difficulties with modeling the background gamma radiation in the galactic plane (see again Thompson,
these proceedings). 

For EGRET, the point spread function is described by the angle $\theta_{68}$, the half-opening angle of
a cone which contains 68 \% of all spark chamber events caused by photons from one point-source.
This angle is weakly energy dependent. It is 5.85$^\circ$ at 100 MeV and 1.7$^\circ$ at 1 GeV.
The expected accuracy of the source location is estimated by   $\theta_{68}/\sqrt{N}$ where
$N$ is the number of detected photons. However, due to systematic errors of the background model also
stronger sources can have location errors of more than 1$^\circ$.

\begin{figure}[ht]
\centerline{\epsfig{file=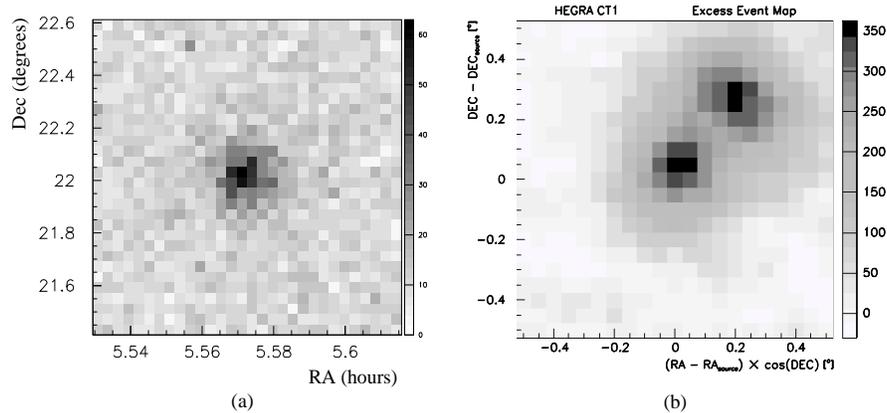, width=\textwidth}}
\caption{ \label{fig-ctloc} Point source location and separation with Cherenkov telescopes: (a)
Excess event map of the region around the Crab Nebula obtained by P\"{u}hlhofer et al. 
(\protect\cite{puehl}) using the HEGRA System of Cherenkov telescopes.
The theoretical accuracy of the location is 0.01$^\circ$. The reconstructed position
deviates from the radio position of the Crab Nebula by 0.01$^\circ$. (b) Excess event map
of a hypothetical double source obtained by superimposing on- and 0.3$^\circ$-off-axis
observations of the Blazar Mkn 501 using the HEGRA CT1 Cherenkov telescope
(Petry \cite{petry}). The point spread function of CT1 is $\theta_{68}=0.1^\circ$ resulting
in an expected location accuracy of 0.01$^\circ$ degrees. Taking into account a systematic 
tracking error of 0.1$^\circ$, the position of Mkn 501 is correctly reconstructed from the on-axis observation. 
More important, the 0.3$^\circ$ angular distance between the two sources, which should be unaffected by 
the tracking error, is also correctly reconstructed within the expected 0.01$^\circ$. }
\end{figure}

Imaging Atmospheric Cherenkov Telescopes (IACTs) are ground-based detectors using the atmosphere
as a tracker and calorimeter. They typically have  point 
spread functions with 
$\theta_{68} < 0.16^\circ$. The number of detected primary gamma-photons is typically $> 100$. 
Hence source locations with arc\-minute accuracy are possible whenever a source can be detected.
This has already been demonstrated by several authors (see the examples in figures \ref{fig-ctloc} a and b).
The possibility of separating two nearby point-sources is demonstrated
by figure \ref{fig-ctloc} (b): Two sources can be separated if their angular distance
is $> 3\theta_{68}$. Cherenkov telescopes therefore seem to be the adequate tool for improving 
the source location of EGRET sources.

With their high photon statistics due to their large collection areas (typically $10^4$m$^2$ 
at the threshold rising up to $10^5$m$^2$ at larger zenith angles and higher energies)
IACTs can also find pulsed emission and resolve fast time structures. This may be
especially helpful since young pulsars (Kaaret \& Cottam \cite{kaaret}) and
Geminga-like radio-quiet pulsars (Yadigaroglu \& Romani \cite{yadi}) are among the candidate sources behind
EGRET UNIDs.

In this article, I would like to explore what the next generation of IACTs with their
improved sensitivity and lower energy threshold can achieve in terms of helping to identify the
EGRET UNIDs. The result will mainly be a list of catalog objects
for which  observations with the new IACTs make sense and for which an identification can be expected.
This defines their possible contribution to the solution of the EGRET UNID puzzle.
As discussed in the next section, I will limit myself to the projects for IACTs with energy 
thresholds below 100 GeV. However, the exercise can easily be repeated for other experiments.

\section{The next-generation Cherenkov telescopes}

Since the middle of the past decade, projects are underway to further improve the
Cherenkov telescope technology and explore the 10-200 GeV regime with ground-based
instruments. At the same time, the new telescopes are aiming to reach much improved 
sensitivity at higher energies.

In this article, I limit myself to those projects which use {\it imaging} atmospheric Cherenkov
telescopes (IACTs), mostly because it has already been demonstrated that this technology
can reach high flux sensitivity and high accuracy in determining source positions.
I would like to mention, however, that other technologies are under development
which may become competitive at some point, especially the non-imaging experiments
using mirror arrays of former solar power plants (CELESTE (e.g. Smith et al. \cite{celeste}), 
GRAAL (e.g. Arqueros et al. \cite{graal}), STACEE (e.g. Oser et al. \cite{stacee}), Solar~II 
(e.g. Zweerink et al. \cite{solar2}) and the
large-field-of-view, large-duty-cycle observatories MILAGRO (e.g. McCullough et al. \cite{milagro})
and ARGO (not yet operational, see e.g. D'Ettore et al. \cite{argo}).

\begin{table}[hbt]
\caption{\label{tab-projects} The four projects for next-generation Imaging Cherenkov Telescopes}
\begin{tabular*}{\textwidth}{@{\extracolsep{\fill}}lccccc}
\hline
Project name & latitude & longitude & altitude & CTs$^a$ & $E_{\mathrm{thr}}(0^\circ)^b$ \\
\hline
CANGAROO III & 31$^\circ$ S & 137$^\circ$ E & 160 m & 4 $\times$ 10 m & 80 GeV \\
HESS I & 23$^\circ$ S & 17$^\circ$ E & 1800 m & 4  $\times$ 13 m & 40 GeV \\
MAGIC I & 29$^\circ$ N &  17$^\circ$ W & 2200 m & 1  $\times$ 17 m & 30 GeV \\
VERITAS & 32$^\circ$ N & 111$^\circ$ W & 1300 m & 7  $\times$ 10 m & 60 GeV \\
\hline
\end{tabular*}
\begin{tablenotes}
$^a$Number of individual telescopes and diameter of their mirror dish.

$^b$Predicted energy threshold for gammas at zenith angle $\vartheta = 0^\circ$.
\end{tablenotes}
\end{table}
\begin{figure}[htb]
\centerline{\epsfig{file=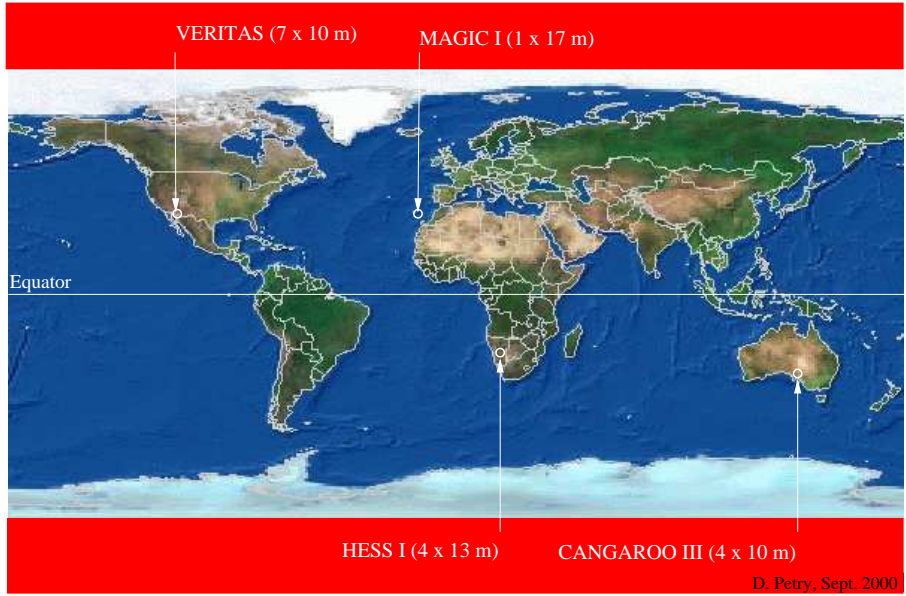, width=\textwidth}}
\caption{ \label{fig-projects} A world map indicating the locations of the four Imaging Atmospheric Cherenkov Telescope
 projects  which may significantly contribute to the identification of EGRET UNIDs. See text for
 references.  }
\end{figure}

The next-generation projects for IACTs which will reach thresholds below 100 GeV are
CANGAROO III (e.g. Mori et al. \cite{mori}), HESS~I (e.g. Hofmann et al. \cite{hess}), 
MAGIC I (e.g. Lorenz et al. \cite{lorenz}) and VERITAS (e.g. Krennrich et al. \cite{veritas}). 
The locations of the sites chosen for these observatories and their estimated minimum energy
thresholds are given in table \ref{tab-projects}.  The locations are also shown in
figure \ref{fig-projects}. The roman numbers behind some of the project names denote
the project phases.

\begin{figure}[hbt]
\centerline{\epsfig{file=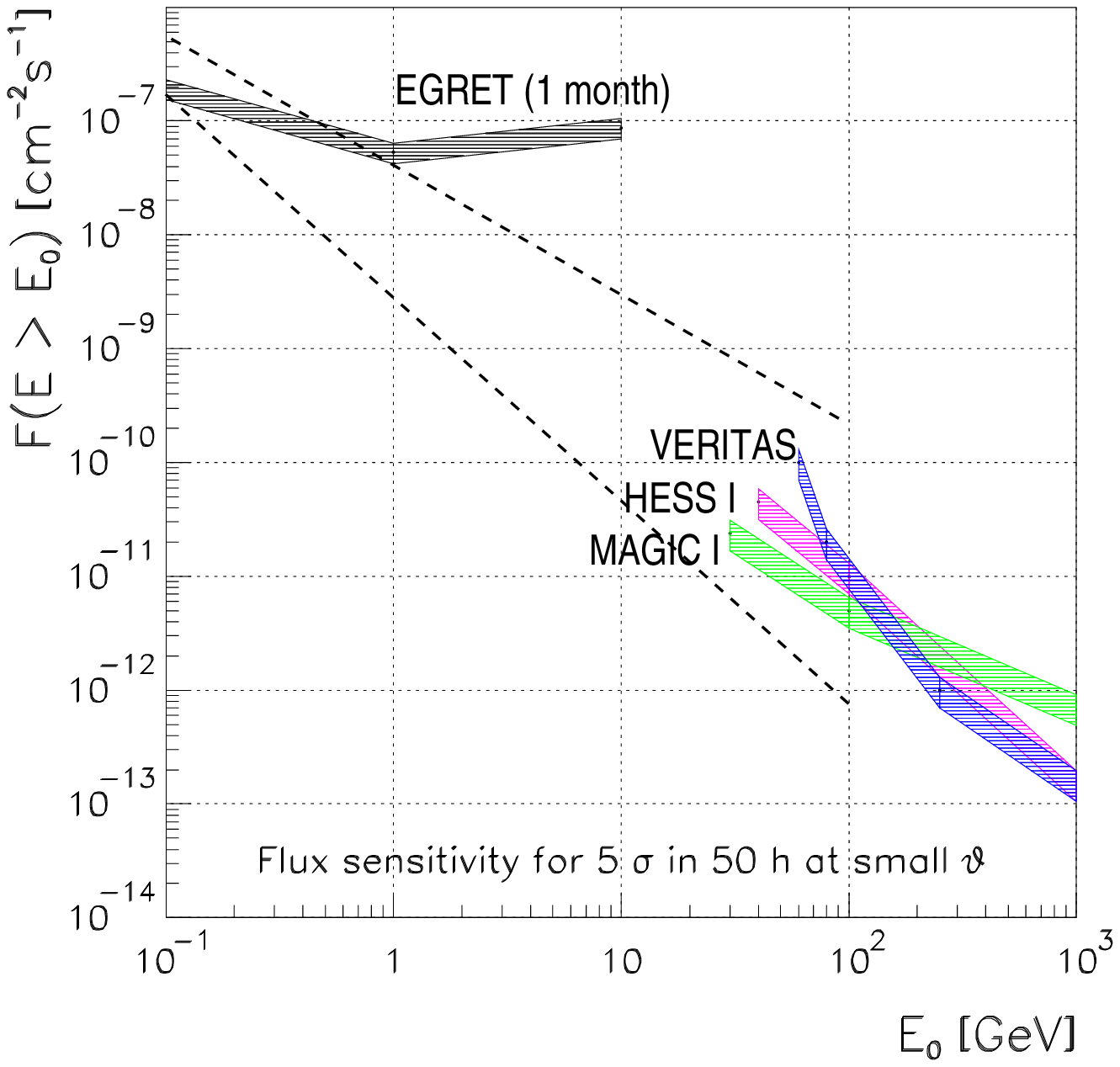, width=0.85\textwidth}}
\caption{ \label{fig-projectsens} The sensitivity of HESS I, MAGIC I and VERITAS projects as
given in Hofmann et al. (\protect\cite{hess}), Barrio et al. (\protect\cite{magic}) and Krennrich et al. 
(\protect\cite{veritas}) respectively. Also shown is the EGRET sensitivity for one month of
observing time. The dashed lines indicate the boundaries of the flux regime in which the central
68 \% of the EGRET unidentified sources can be found if their 100 MeV fluxes are extrapolated
to 100 GeV using the spectrum measured by EGRET. }
\end{figure}

All four experiments except CANGAROO III have published detailed Monte Carlo studies
and presented a sensitivity curve. Figure \ref{fig-projectsens} shows these sensitivity curves 
in the relevant energy range up to 1 TeV. 
CANGAROO III will reach a performance roughly similar to VERITAS (Mori \cite{mori2}).
The sensitivity curves show the integral gamma-ray flux which the
corresponding instrument can detect above the energy threshold $E_0$
with 20 \% statistical error within an observation time
of 50 h at small zenith angles. Also shown is the EGRET flux sensitivity for an observation
time of 1 month taking into account the fact that the field of view and the duty cycle
of EGRET is larger than that of an IACT.

Furthermore, figure \ref{fig-projectsens} indicates in which flux regime the bulk part of
the EGRET unidentified objects (UNIDs) would be found if their spectrum would continue
as a straight power law from the 100 MeV, at which it is measured by EGRET, to 100 GeV.
Plotting the distribution of the logarithms of the extrapolated fluxes at 100 GeV 
gives a nicely symmetric, nearly Gaussian distribution with 
an average of -10.79 corresponding to $1.6 \times 10^{-11}$ cm$^{-2}$s$^{-1}$ and a standard
deviation of 1.2. The dashed lines in figure \ref{fig-projectsens} outline the flux regime
one standard deviation above and below this average.

\section{Observability criteria}

\label{sec-crit}

Figure \ref{fig-projectsens} is misleading as far as the actual observability of
EGRET UNIDs by the IACTs under consideration is concerned. There are four aspects
which have to be taken into account
in assembling an observation agenda. These are discussed in the following.

\subsection{Spectral steepening}

 The most obvious problem is that most EGRET sources will show spectral breaks
 above 100 MeV and the naive extrapolation of the power-law measured at around 100 MeV 
 may be orders of magnitude above the real flux at the IACT threshold.

 In order to predict observability more reliably, a certain spectral steepening 
 has to be assumed. At the same time, no bias should be introduced towards any source
 class.  
 As a first general approach, I will therefore assume a spectral steepening similar to
 that of the Crab Nebula which is the best studied high energy gamma source.
 This is also adequate since most of the EGRET UNIDs are expected to be galactic
 such that cut-offs due to the absorption in the intergalactic infrared background
 should not be assumed a priori.

 The differential spectral index $\gamma_{0.1}$ of the Crab Nebula at 0.1 GeV is 
given in the Third EGRET Catalog as $\gamma_{0.1} = 2.19 \pm 0.02$ while it
is measured by the Whipple telescope at 500 GeV to be $\gamma_{500} = 2.49 \pm 0.06 \pm 0.04$
(Hillas et al. \cite{hillas}). However, since the EGRET energy range is situated at
the transition of the synchrotron component to the Inverse-Compton (IC) component of
the Crab spectrum, the index 2.19 is actually an interpolation between the
steep synchrotron spectrum and the beginning flat IC spectrum (de Jager et al. \cite{dejager}). Above 1 GeV, the
IC component is dominant. Hillas et al. (\cite{hillas}) show that the steepening
of the IC component towards higher energies can be described by an increase
in spectral index of 0.15 per decade of energy.

Applied to the extrapolation of the EGRET UNID spectra, this would mean that
the integral flux above the energy $x$ $[$GeV$]$, $10 < x < 1000$ can roughly be expected to be
\begin{equation} \label{equ-extraflux}
\begin{array}{rcl}
   F ( E > x [\mathrm{GeV}]) & = & F_0 \cdot 10^{-\alpha} \cdot 10^{-(\alpha+0.15)} \cdot (\frac{x}{10})^{-(\alpha+0.30)}\\
				  & = & F_0 \cdot 10^{-(\alpha - 0.15)} \cdot x^{-(\alpha+0.30)}\\
				& & \mathrm{ for}\ \  10 < x < 100 \\
  \\
   F ( E > x [\mathrm{GeV}]) & = & F_0 \cdot 10^{-\alpha} \cdot 10^{-(\alpha+0.15)} \cdot 10^{-(\alpha+0.30)} \cdot (\frac{x}{100})^{-(\alpha+0.45)}\\
				  & = & F_0 \cdot 10^{-(\alpha - 0.45)} \cdot x^{-(\alpha+0.45)}\\
				& & \mathrm{ for}\ \  100 < x < 1000 \\
\end{array}
\end{equation}
where $F_0$ is the integral flux above 0.1 GeV of a given source taken from the Third EGRET Catalog 
and $\alpha$ is the integral spectral index derived 
by subtracting 1.0 from the differential spectral index stated in the same catalog.
E.g. for $\alpha = 1.0$, the steepening leads to an extrapolated integral flux at 100 GeV
which is smaller by a factor 2.8 than the extrapolation without steepening.

\subsection{Sky access limits}

Due to the decaying optical quality of the atmosphere towards larger zenith angles 
(water vapor, aerosols, background light from towns etc.), Cherenkov telescopes typically cannot
observe at zenith angles much larger than 70$^\circ$. And even if they can,
they will not give priority to observations of objects which can only be
observed at zenith angles larger than this.

The zenith angle $\vartheta$ at the upper culmination of an astronomical object depends
on the the latitude $\phi$ of the observatory and the declination DEC of the object
according to $ \vartheta = | \phi - \mathrm{DEC} | $. Hence the condition
\begin{equation}
 | \phi - \mathrm{DEC} | < 70^\circ
\end{equation}
has to be imposed in the selection of observable objects. In order to make sure that
the objects can be observed for sufficient time at $\vartheta<70^\circ$, the
permitted declination-range has to be narrowed down to $| \phi - \mathrm{DEC} | < 60^\circ$
for objects which are not circumpolar.
For the four projects under consideration these conditions mean

\begin{equation} \label{equ-decconstraint}
\begin{array}{lrcl}
\mathrm{CANGAROO\ III} &    & \mathrm{DEC} &  < +29^\circ\\
\mathrm{HESS\ I}       &    & \mathrm{DEC} &  < +37^\circ\\
\mathrm{MAGIC\ I}      & -31^\circ <  & \mathrm{DEC} & \\
\mathrm{VERITAS}      & -28^\circ <  & \mathrm{DEC} & \\
\end{array}
\end{equation}

\begin{figure}[hbt]
\centerline{\epsfig{file=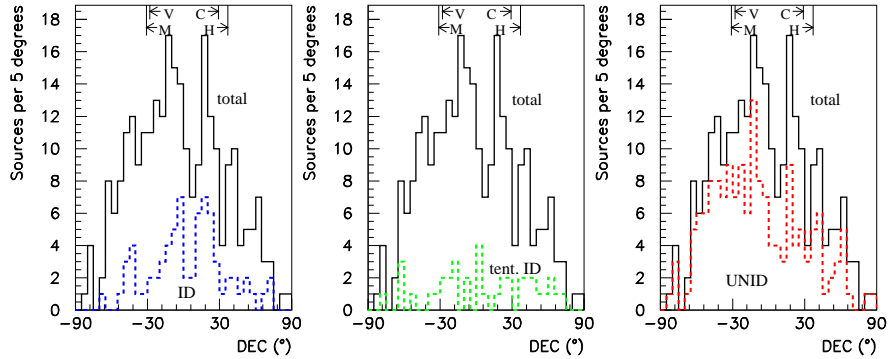, width=\textwidth}}
\caption{ \label{fig-decdist} The distributions of the declination (DEC)
of the objects in the third EGRET catalog. Left: the identified objects.
Middle: the tentatively identified objects. Right: the unidentified objects.
On all plots the distribution for the total number of sources is superimposed.
At the top of each plot, arrows indicate the sky access limits of CANGAROO III (C),
HESS I (H), MAGIC I (M) and VERITAS (V).}
\end{figure}

\noindent
Figure \ref{fig-decdist} shows the DEC distributions for the identified,
tentatively identified and unidentified EGRET objects in the third catalog.
The figure also indicates the DEC constraints for the four IACT projects.
Due to the position of the galactic plane, the projects on the southern hemisphere 
clearly have better access to EGRET UNIDs. The distributions of identified and
tentatively identified objects are nearly symmetrical.
	
\subsection{Zenith-angle dependence of energy threshold and flux sensitivity}

Figure \ref{fig-projectsens} shows the sensitivity of the various IACT projects
for small zenith angles ($\vartheta < 20^\circ$) just like it shows the EGRET sensitivity 
for on-axis observations. Both effective collection area $A_{\mathrm{eff}}$
and energy threshold $E_{\mathrm{thr}}$ of an IACT are to a good approximation
proportional to $\cos^{-2}(\vartheta)$ as explained in the following.

The area perpendicular to the optical axis 
illuminated by the Che\-ren\-kov light at the position of the telescope
 is proportional to the square
of the distance $d$ to the shower maximum. $d$ grows with $\vartheta$ as $d \propto 1/\cos(\vartheta)$.
The energy threshold $E_{\mathrm{thr}}$
 of a Cherenkov telescope is, for a given zenith angle, to a good approximation equivalent 
to a Cherenkov photon density threshold
$\rho_{\mathrm{thr}}$ at the position of the telescope. $\rho_{\mathrm{thr}}$ is an
instrumental constant determined by the trigger condition of the data acquisition system.
The Cherenkov photon density $\rho$ at the position of the telescope is a function of
the primary Energy $E$ and of the zenith angle $\vartheta$. The dependence on the
impact parameter can be neglected since we average over it. For a given primary energy,
$\rho$ decreases inversely proportional to the illuminated area, i.e. $\rho(E, \vartheta) \propto \cos^2(\vartheta)$.
At least for electromagnetic showers,
the atmosphere is (again to a good approximation) a linear calorimeter yielding a Cherenkov 
photon density directly proportional to the primary energy of the incident gamma photon,
i.e. for a fixed $\vartheta$  we  have $\rho(E, \vartheta) \propto E$.
Hence, $\rho(E, \vartheta) \propto E\cdot\cos^2(\vartheta)$. It follows that
to satisfy the trigger condition $\rho(E,\vartheta) > \rho_{\mathrm{thr}}$, $E$ has to
increase with $\vartheta$ as $\cos^{-2}(\vartheta)$.
\begin{equation} \label{equ-ethr}
\begin{array}{rcl}
    A_{\mathrm{eff}}(\vartheta) & = & A_{\mathrm{eff}}(0^\circ) \cdot \cos^{-2}(\vartheta)\\
    E_{\mathrm{thr}}(\vartheta) & = & E_{\mathrm{thr}}(0^\circ) \cdot \cos^{-2}(\vartheta)\\
\end{array}
\end{equation}
The values for $E_{\mathrm{thr}}(0^\circ)$ are shown in table \ref{tab-projects}.

An increase in effective collection area for gammas is accompanied by a proportional increase
in hadronic background rate. The gain in flux sensitivity is therefore only the square-root
of the gain in area:
\begin{equation}
   F_{5\sigma}( E_{\mathrm{thr}}(\vartheta), \vartheta ) = 
	 F_{5\sigma}( E_{\mathrm{thr}}(0^\circ), 0^\circ ) \cdot \cos^{-1}(\vartheta)
\end{equation}
where $F_{5\sigma}( E_{\mathrm{thr}})$ is the integral flux above energy threshold $E_{\mathrm{thr}}$
which results in a $5 \sigma$ detection after 50~h of observation time.

\subsection{Required observation time}

Given an integral flux $F(E_{\mathrm{thr}})$ above an energy threshold $E_{\mathrm{thr}}$
and a flux sensitivity expressed as $F_{5\sigma}( E_{\mathrm{thr}})$ as defined in the
previous section, and assuming that the flux is so small that the signal is background
dominated (the conservative assumption), we can calculate the observation time $T_{5\sigma}$
which is necessary to detect the flux with $5 \sigma$ significance:
\begin{equation} \label{equ-obstime}
   T_{5\sigma}(E_{\mathrm{thr}}) = \left( \frac{F(E_{\mathrm{thr}})}{F_{5\sigma}( E_{\mathrm{thr}})} \right) ^{-2} \cdot 50 \mathrm{h}
\end{equation}

Cherenkov telescopes typically have a duty cycle of about 10 \% due to the fact that they can only observe 
during good weather and at night (moon-less night for maximum sensitivity).
This results typically in 1000~h of available observation time per year.
For an individual source, rarely more than 100~h can be obtained.

\section{The agenda}

Based on the observability criteria developed above, UNIDs can now be selected for
an observation agenda. In the following, those sources are preselected which
may be detected in less than 50 h of observation time. These are shown in
tables \ref{tab-agenda-canga},
\ref{tab-agenda-hess},  \ref{tab-agenda-magic}, and \ref{tab-agenda-veritas} for
further reference. The prime candidates are then selected requiring that a detection
is possible even if the spectrum at 0.1 GeV is steeper than the nominal value
by one standard deviation. The prime candidates are summarized in table \ref{tab-agenda}.
A discussion follows in section \ref{sec-disc}.

\subsection{Preselection}

Using the formulae derived in section \ref{sec-crit}, tables \ref{tab-agenda-canga},
\ref{tab-agenda-hess},  \ref{tab-agenda-magic}, and \ref{tab-agenda-veritas}
are compiled. All four tables have the same format.
The tables are generated as follows:
For each non-identified source of the third catalog and for each of the
four IACT projects, the minimum zenith angle of observation $\vartheta_{\mathrm{min}}$ is calculated
using
\begin{equation} \label{equ-thetamin}
   \vartheta_{\mathrm{min}} = | \phi - DEC |
\end{equation}
and the corresponding minimum energy threshold $E_{\mathrm{thr}}(\vartheta_{\mathrm{min}})$
is determined according to equation \ref{equ-ethr}. 
$\vartheta_{\mathrm{min}}$ and the corresponding threshold are shown in the tables. 
If a source
is not observable according to equation \ref{equ-decconstraint}, it is omitted.
Otherwise, the threshold is inserted into equation \ref{equ-extraflux} to determine the
extrapolated flux $F(E_{\mathrm{thr}})$. From this flux, the ne\-cessary observation
time $T_{5\sigma}$ is determined using equation \ref{equ-obstime}. Flux and observation
time together with the spectral index $\alpha(E_{\mathrm{thr}})$ which the extrapolated 
spectrum would have at the IACT threshold,
are also shown in the table. If $T_{5\sigma}$ is more than 50~h,
the source is omitted. 

In order to take into account the error of the EGRET spectral measurement
and to explore a realistic worst case, the procedure is repeated assuming
an initial spectral index at 0.1 GeV which is larger by one standard deviation
than the nominal value. The resulting modified values for $F(E_{\mathrm{thr}})$,
$\alpha(E_{\mathrm{thr}})$ and $T_{5\sigma}$ are given in brackets after
their nominal values. The modified value of $T_{5\sigma}$ is named $T_{5\sigma}'$
in the following.

For the case of VERITAS with its steep dependence of $F_{5\sigma}$ on $E$ below 80 GeV (see
figure \ref{fig-projectsens}), it turns out that for all sources the observation
time can be reduced if the minimum threshold is raised to 80 GeV before entering
the calculation. The table contains the values for this case.

After $T_{5\sigma}$ is calculated, the number of detected primary gamma photons
is calculated conservatively assuming an effective collection area of $10^4$~m$^2$.
If this number is smaller than 100, the value of $T_{5\sigma}$ is increased
such that the number becomes equal to 100. This ensures that the source
location can be determined with sufficiently high accuracy to improve
the EGRET measurement. The values are marked if this correction was necessary.

\subsection{Prime candidates}

Table \ref{tab-agenda} shows those sources from tables  \ref{tab-agenda-canga},
\ref{tab-agenda-hess},  \ref{tab-agenda-magic}, and \ref{tab-agenda-veritas} which satisfy
the additional restriction that the ``modestly worst case'' observation time $T_{5\sigma}'$
(see the previous section) is less than 50 h. These sources can be regarded as the
prime candidates for allocating the valuable IACT observation time since only if a detection
is possible, the EGRET location measurement can be improved. An upper limit on the
flux value is only valuable if the observed object has been identified and models can be
constrained.

\section{Discussion}

\label{sec-disc}

The observability  tables \ref{tab-agenda-canga},
\ref{tab-agenda-hess},  \ref{tab-agenda-magic}, and \ref{tab-agenda-veritas}
and the selection of prime candidates in table \ref{tab-agenda}
show that the next-generation IACTs may make a major contribution to the
identification of EGRET UNIDs. Altogether 78 of them may be in the reach of
the IACT sensitivity. Under the modest assumptions of spectral steepening made here,
many sources may be detectable within few hours of observation time.

Among the 78 sources are 20 which have a tentative identification 
(given in the Third EGRET Calalog unless otherwise specified): 

\noindent
\begin{sloppypar}
 0010+7309 (SNR CTA1),
 0241+6103 (LS I +61$^\circ$303), 
 0617+2238 (SNR IC443),
 0808+4844 (QSO 0809+483),
 0812-0646 (PKS 0805-077),
 0827-4247 (SNR Pup A)
 0917+4427 (QSO 0917+449), 
 1009+4855 (QSO 1011+496), 
 1323+2200 (EQ 1324+224), 
 1410-6147 (SNR 312.4-00.4),
 1627-2419 (rho Oph),
 1800-2338 (SNR  W 28),
 1854-1514 (LS5039, a microquasar suggested by Paredes et al. (\cite{paredes})),  
 1856+0114 (SNR W 44), 
 1903+0550 (SNR 040.4-00.5), 
 2020+4017 (SNR $\gamma$Cyg), 
 2100+6012 (4C 59.33), 
 2206+6602 (TXS 2206+650),
 2255+1943 (QSO 2250+1926), and
 2352+3752 (QSO 2346+385). 
\end{sloppypar}

Some of these have already (unsuccessfully) been observed
by present IACTs. They come as no surprise and will be
included in observation schedules in any case since the
known position of the possible counterpart makes the
observation especially promissing. 

Among the 38 prime candidates shown in table \ref{tab-agenda}, only 11
of the tentatively identified objects are present, mostly SNR candidates.
Observers will face the additional problem here that these sources may be
extended in angular size which may increase the necessary observation time.
This is also true for all other UNIDs. 

Increases in the required
observation time will also arise in those few cases where $\theta_{95}$, the radius of
the 3EG position probability map, is near the radius of the IACT trigger 
field-of-view (typically $> 1^\circ$) and the true source position is accidentally 
on the boundary of the map. But most of the 78 selected UNIDs have a $\theta_{95}$
well below $1^\circ$ (see tables \ref{tab-agenda-canga},
\ref{tab-agenda-hess},  \ref{tab-agenda-magic}, and \ref{tab-agenda-veritas}). 

A decrease in necessary observation time may be possible in case a candidate counterpart-pulsar 
is known. The IACT sensitivity can be greatly increased by introducing the
known pulsar ephemeris into the analysis provided the emission is still pulsed at
the IACT's threshold. 

A nice opportunity to cross-calibrate all four IACT observatories are the prime candidate
objects 1837-0606 and 1856+0114 (SNR W44?) which are observable from all four locations
at similar zenith angles (see table \ref{tab-agenda}).

In summary, the EGRET UNIDs are a promissing field for the application of the
next-generation IACTs. From the start of operations of the individual observatory
it will probably take between 2 and 4 years until the between 16 and 27 prime candidates
have been observed and analysed. This estimation takes into account that
not the whole observation time will be dedicated to EGRET UNIDs.
However, if the observatories cooperate and split the candidate list into four parts
of 9 to 10 sources each, a coverage of all 38 prime candidates could be reached within one year since 
each observatory would only have to invest up to $\approx$ 500~h of observation time.
The attempt to cover the remaining 40 non-prime candidates could equally be shared.

\begin{table}[hbt]
\caption{\label{tab-agenda-canga} Observability of unidentified and tentatively identified EGRET sources for project CANGAROO III}
{\small
\begin{tabular}{l@{\hspace{0.15cm}}c@{\hspace{0.15cm}}c@{\hspace{0.15cm}}c@{\hspace{0.15cm}}c@{\hspace{0.15cm}}c@{\hspace{0.15cm}}c@{\hspace{0.15cm}}c}
\hline
\multicolumn{3}{c}{3EG} & \multicolumn{5}{|c}{ CANGAROO III observability} \\
object  & \multicolumn{1}{c}{$\theta_{95}$$^b$}  
	& \multicolumn{1}{c}{$\alpha$$^c$}  
	& $\vartheta_{\mathrm{min}}$$^d$ 
	& \multicolumn{1}{c}{$E_{\mathrm{thr}}$$^e$} 
	& \multicolumn{1}{c}{$F(E_{\mathrm{thr}})$$^f$} 
	& \multicolumn{1}{c}{$\alpha(E_{\mathrm{thr}})$$^g$} 
	& \multicolumn{1}{c}{$T_{5\sigma}$$^h$}  \\ 
name    &\multicolumn{1}{c}{$[^\circ]$} 
	&  
	& \multicolumn{1}{c}{$[^\circ]$} 
	& \multicolumn{1}{c}{$[$GeV$]$} 
	& \multicolumn{1}{c}{$[$cm$^{-2}$s$^{-1}$$]$} 
	&  
	& \multicolumn{1}{c}{$[$h$]$}  \\
\hline
 0215+1123 &   1.06 &  $1.03\pm0.62$ &   42 &    $147$ &    2.95E-11(3.21E-13) & 1.48(2.10) &  $ 42$$(>500)$ \\ 
 0616-3310 &   0.63 &  $1.11\pm0.24$ &    2 &    $ 80$ &    2.86E-11(5.74E-12) & 1.41(1.65) &  $ 25$$(>500)$ \\ 
 0617+2238$^a$ &   0.13 &  $1.01\pm0.06$ &   54 &    $228$ &    5.13E-11(3.22E-11) & 1.46(1.52) &  $ 22$$( 55)$ \\ 
 0631+0642 &   0.46 &  $1.06\pm0.15$ &   38 &    $128$ &    4.12E-11(1.41E-11) & 1.51(1.66) &  $ 19$$(161)$ \\ 
 0634+0521 &   0.67 &  $1.03\pm0.26$ &   36 &    $123$ &    5.39E-11(8.47E-12) & 1.48(1.74) &  $ 11$$(430)$ \\ 
 0706-3837 &   0.90 &  $1.30\pm0.43$ &    8 &    $ 81$ &    3.23E-11(1.81E-12) & 1.60(2.03) &  $ 20$$(>500)$ \\ 
 0747-3412 &   0.70 &  $1.22\pm0.30$ &    3 &    $ 80$ &    3.09E-11(4.16E-12) & 1.52(1.82) &  $ 21$$(>500)$ \\ 
 0824-4610 &   0.61 &  $1.36\pm0.07$ &   15 &    $ 86$ &    2.43E-11(1.51E-11) & 1.66(1.73) &  $ 36$$( 94)$ \\ 
 0827-4247$^a$ &   0.77 &  $1.10\pm0.12$ &   12 &    $ 83$ &    9.75E-11(4.35E-11) & 1.40(1.52) &  $2.85^\ddag$$( 11)$ \\ 
 0841-4356 &   0.52 &  $1.15\pm0.09$ &   13 &    $ 84$ &    1.14E-10(6.23E-11) & 1.45(1.54) &  $2.43^\ddag$$(5.43)$ \\ 
 0848-4429 &   0.62 &  $1.05\pm0.16$ &   14 &    $ 85$ &    2.31E-10(7.87E-11) & 1.35(1.51) &  $1.20^\ddag$$(3.53^\ddag)$ \\ 
 1014-5705 &   0.67 &  $1.23\pm0.20$ &   26 &    $ 99$ &    4.73E-11(1.19E-11) & 1.53(1.73) &  $ 11$$(175)$ \\ 
 1027-5817 &   0.37$^*$ &  $0.94\pm0.09$ &   27 &    $101$ &    3.48E-10(1.87E-10) & 1.39(1.48) &  $0.80^\ddag$$(1.49^\ddag)$ \\ 
 1048-5840 &   0.17 &  $0.97\pm0.09$ &   28 &    $102$ &    2.62E-10(1.41E-10) & 1.42(1.51) &  $1.06^\ddag$$(1.98^\ddag)$ \\ 
 1234-1318 &   0.76 &  $1.09\pm0.24$ &   18 &    $ 88$ &    4.91E-11(9.63E-12) & 1.39(1.63) &  $9.16$$(237)$ \\ 
 1323+2200$^a$ &   0.47 &  $0.86\pm0.35$ &   53 &    $221$ &    5.98E-11(4.04E-12) & 1.31(1.66) &  $ 15$$(>500)$ \\ 
 1410-6147$^a$ &   0.36 &  $1.12\pm0.14$ &   31 &    $108$ &    8.78E-11(3.30E-11) & 1.57(1.71) &  $3.51$$( 25)$ \\ 
 1420-6038 &   0.32 &  $1.02\pm0.14$ &   30 &    $106$ &    2.10E-10(7.91E-11) & 1.47(1.61) &  $1.33^\ddag$$(4.24)$ \\ 
 1627-2419$^a$ &   0.65 &  $1.21\pm0.27$ &    7 &    $ 81$ &    2.67E-11(4.38E-12) & 1.51(1.78) &  $ 28$$(>500)$ \\ 
 1631-4033 &   0.89 &  $1.25\pm0.27$ &   10 &    $ 82$ &    2.24E-11(3.65E-12) & 1.55(1.82) &  $ 41$$(>500)$ \\ 
 1655-4554 &   0.66 &  $1.19\pm0.24$ &   15 &    $ 86$ &    4.63E-11(9.16E-12) & 1.49(1.73) &  $9.99$$(255)$ \\ 
 1704-4732 &   0.66 &  $0.86\pm0.33$ &   17 &    $ 87$ &    1.29E-09(1.38E-10) & 1.16(1.49) &  $0.22^\ddag$$(2.01^\ddag)$ \\ 
 1714-3857 &   0.51 &  $1.30\pm0.20$ &    8 &    $ 82$ &    2.70E-11(7.06E-12) & 1.60(1.80) &  $ 28$$(409)$ \\ 
 1736-2908 &   0.62 &  $1.18\pm0.12$ &    2 &    $ 80$ &    7.32E-11(3.28E-11) & 1.48(1.60) &  $3.79^\ddag$$( 19)$ \\ 
 1741-2050 &   0.63 &  $1.25\pm0.12$ &   10 &    $ 83$ &    2.05E-11(9.14E-12) & 1.55(1.67) &  $ 49$$(247)$ \\ 
 1744-3011 &   0.32 &  $1.17\pm0.08$ &    1 &    $ 80$ &    9.72E-11(5.70E-11) & 1.47(1.55) &  $2.86^\ddag$$(6.17)$ \\ 
 1746-2851 &   0.13 &  $0.70\pm0.07$ &    2 &    $ 80$ &    4.22E-09(2.64E-09) & 1.00(1.07) &  $0.07^\ddag$$(0.11^\ddag)$ \\ 
 1800-2338$^a$ &   0.32 &  $1.10\pm0.10$ &    7 &    $ 81$ &    1.46E-10(7.45E-11) & 1.40(1.50) &  $1.91^\ddag$$(3.73^\ddag)$ \\ 
 1809-2328 &   0.16 &  $1.06\pm0.08$ &    8 &    $ 81$ &    1.29E-10(7.57E-11) & 1.36(1.44) &  $2.15^\ddag$$(3.67^\ddag)$ \\ 
 1812-1316 &   0.39 &  $1.29\pm0.11$ &   18 &    $ 88$ &    2.65E-11(1.26E-11) & 1.59(1.70) &  $ 31$$(139)$ \\ 
 1824-1514 &   0.52 &  $1.19\pm0.18$ &   16 &    $ 86$ &    4.18E-11(1.24E-11) & 1.49(1.67) &  $ 12$$(141)$ \\ 
 1826-1302 &   0.46 &  $1.00\pm0.11$ &   18 &    $ 88$ &    2.78E-10(1.32E-10) & 1.30(1.41) &  $1.00^\ddag$$(2.11^\ddag)$ \\ 
 1837-0606 &   0.19 &  $0.82\pm0.14$ &   25 &    $ 97$ &    6.30E-10(2.40E-10) & 1.12(1.26) &  $0.44^\ddag$$(1.16^\ddag)$ \\ 
 1850-2652 &   1.00 &  $1.29\pm0.45$ &    4 &    $ 80$ &    6.08E-11(2.99E-12) & 1.59(2.04) &  $5.44$$(>500)$ \\ 
 1856+0114$^a$ &   0.19 &  $0.93\pm0.10$ &   32 &    $112$ &    3.33E-10(1.65E-10) & 1.38(1.48) &  $0.83^\ddag$$(1.68^\ddag)$ \\ 
 1928+1733 &   0.75 &  $1.23\pm0.32$ &   49 &    $183$ &    4.13E-11(3.74E-12) & 1.68(2.00) &  $ 27$$(>500)$ \\ 
\hline
\end{tabular}
\begin{tablenotes}
For definitions see the end of table \ref{tab-agenda-veritas}.
\end{tablenotes}

}
\end{table}

\begin{table}[hbt]
\caption{\label{tab-agenda-hess} Observability of unidentified and tentatively identified EGRET sources for project HESS I}
{\small
\begin{tabular}{l@{\hspace{0.15cm}}c@{\hspace{0.15cm}}c@{\hspace{0.15cm}}c@{\hspace{0.15cm}}c@{\hspace{0.15cm}}c@{\hspace{0.15cm}}c@{\hspace{0.15cm}}c}
\hline
\multicolumn{3}{c}{3EG} & \multicolumn{5}{|c}{ HESS I observability} \\
object  & \multicolumn{1}{c}{$\theta_{95}$$^b$}  
	& \multicolumn{1}{c}{$\alpha$$^c$}  
	& $\vartheta_{\mathrm{min}}$$^d$ 
	& \multicolumn{1}{c}{$E_{\mathrm{thr}}$$^e$} 
	& \multicolumn{1}{c}{$F(E_{\mathrm{thr}})$$^f$} 
	& \multicolumn{1}{c}{$\alpha(E_{\mathrm{thr}})$$^g$} 
	& \multicolumn{1}{c}{$T_{5\sigma}$$^h$}  \\ 
name    &\multicolumn{1}{c}{$[^\circ]$} 
	&  
	& \multicolumn{1}{c}{$[^\circ]$} 
	& \multicolumn{1}{c}{$[$GeV$]$} 
	& \multicolumn{1}{c}{$[$cm$^{-2}$s$^{-1}$$]$} 
	&  
	& \multicolumn{1}{c}{$[$h$]$}  \\
\hline
 0215+1123 &   1.06 &  $1.03\pm0.62$ &   34 &    $ 59$ &    1.05E-10(2.02E-12) & 1.33(1.95) &  $ 13$$(>500)$ \\ 
 0616-3310 &   0.63 &  $1.11\pm0.24$ &   10 &    $ 41$ &    7.28E-11(1.72E-11) & 1.41(1.65) &  $ 20$$(355)$ \\ 
 0617+2238$^a$ &   0.13 &  $1.01\pm0.06$ &   46 &    $ 82$ &    2.21E-10(1.48E-10) & 1.31(1.37) &  $4.22$$(9.44)$ \\ 
 0631+0642 &   0.46 &  $1.06\pm0.15$ &   30 &    $ 53$ &    1.42E-10(5.53E-11) & 1.36(1.51) &  $6.69$$( 44)$ \\ 
 0634+0521 &   0.67 &  $1.03\pm0.26$ &   28 &    $ 52$ &    1.77E-10(3.49E-11) & 1.33(1.59) &  $4.17$$(107)$ \\ 
 0706-3837 &   0.90 &  $1.30\pm0.43$ &   16 &    $ 43$ &    8.92E-11(6.57E-12) & 1.60(2.03) &  $ 14$$(>500)$ \\ 
 0747-3412 &   0.70 &  $1.22\pm0.30$ &   11 &    $ 42$ &    8.40E-11(1.38E-11) & 1.52(1.82) &  $ 15$$(>500)$ \\ 
 0824-4610 &   0.61 &  $1.36\pm0.07$ &   23 &    $ 47$ &    6.53E-11(4.24E-11) & 1.66(1.73) &  $ 28$$( 67)$ \\ 
 0827-4247$^a$ &   0.77 &  $1.10\pm0.12$ &   20 &    $ 45$ &    2.30E-10(1.11E-10) & 1.40(1.52) &  $2.15$$(9.34)$ \\ 
 0841-4356 &   0.52 &  $1.15\pm0.09$ &   21 &    $ 46$ &    2.76E-10(1.59E-10) & 1.45(1.54) &  $1.53$$(4.61)$ \\ 
 0848-4429 &   0.62 &  $1.05\pm0.16$ &   22 &    $ 46$ &    5.24E-10(1.96E-10) & 1.35(1.51) &  $0.53^\ddag$$(3.04)$ \\ 
 0859-4257 &   0.64 &  $1.32\pm0.20$ &   20 &    $ 45$ &    5.09E-11(1.50E-11) & 1.62(1.82) &  $ 44$$(>500)$ \\ 
 1014-5705 &   0.67 &  $1.23\pm0.20$ &   34 &    $ 58$ &    1.07E-10(2.98E-11) & 1.53(1.73) &  $ 13$$(166)$ \\ 
 1027-5817 &   0.37$^*$ &  $0.94\pm0.09$ &   35 &    $ 60$ &    6.67E-10(3.75E-10) & 1.24(1.33) &  $0.42^\ddag$$(1.08)$ \\ 
 1048-5840 &   0.17 &  $0.97\pm0.09$ &   36 &    $ 61$ &    5.09E-10(2.86E-10) & 1.27(1.36) &  $0.59$$(1.87)$ \\ 
 1234-1318 &   0.76 &  $1.09\pm0.24$ &   10 &    $ 41$ &    1.41E-10(3.33E-11) & 1.39(1.63) &  $5.22$$( 94)$ \\ 
 1323+2200$^a$ &   0.47 &  $0.86\pm0.35$ &   45 &    $ 80$ &    2.19E-10(2.11E-11) & 1.16(1.51) &  $4.23$$(456)$ \\ 
 1410-6147$^a$ &   0.36 &  $1.12\pm0.14$ &   39 &    $ 66$ &    1.80E-10(7.28E-11) & 1.42(1.56) &  $5.11$$( 31)$ \\ 
 1420-6038 &   0.32 &  $1.02\pm0.14$ &   38 &    $ 64$ &    4.13E-10(1.67E-10) & 1.32(1.46) &  $0.95$$(5.78)$ \\ 
 1627-2419$^a$ &   0.65 &  $1.21\pm0.27$ &    1 &    $ 40$ &    7.76E-11(1.54E-11) & 1.51(1.78) &  $ 17$$(428)$ \\ 
 1631-4033 &   0.89 &  $1.25\pm0.27$ &   18 &    $ 44$ &    5.90E-11(1.14E-11) & 1.55(1.82) &  $ 32$$(>500)$ \\ 
 1634-1434 &   0.49$^*$ &  $1.15\pm0.23$ &    8 &    $ 41$ &    5.30E-11(1.33E-11) & 1.45(1.68) &  $ 37$$(>500)$ \\ 
 1655-4554 &   0.66 &  $1.19\pm0.24$ &   23 &    $ 47$ &    1.13E-10(2.57E-11) & 1.49(1.73) &  $9.38$$(180)$ \\ 
 1704-4732 &   0.66 &  $0.86\pm0.33$ &   25 &    $ 48$ &    2.55E-09(3.32E-10) & 1.16(1.49) &  $0.11^\ddag$$(1.11)$ \\ 
 1714-3857 &   0.51 &  $1.30\pm0.20$ &   16 &    $ 43$ &    7.44E-11(2.21E-11) & 1.60(1.80) &  $ 20$$(224)$ \\ 
 1717-2737 &   0.64 &  $1.23\pm0.15$ &    5 &    $ 40$ &    5.19E-11(2.11E-11) & 1.53(1.68) &  $ 38$$(229)$ \\ 
 1719-0430 &   0.44 &  $1.20\pm0.24$ &   18 &    $ 44$ &    4.87E-11(1.13E-11) & 1.50(1.74) &  $ 47$$(>500)$ \\ 
 1736-2908 &   0.62 &  $1.18\pm0.12$ &    6 &    $ 40$ &    2.01E-10(9.78E-11) & 1.48(1.60) &  $2.53$$( 11)$ \\ 
 1741-2050 &   0.63 &  $1.25\pm0.12$ &    2 &    $ 40$ &    6.28E-11(3.06E-11) & 1.55(1.67) &  $ 26$$(108)$ \\ 
 1744-3011 &   0.32 &  $1.17\pm0.08$ &    7 &    $ 41$ &    2.63E-10(1.63E-10) & 1.47(1.55) &  $1.48$$(3.88)$ \\ 
 1746-2851 &   0.13 &  $0.70\pm0.07$ &    6 &    $ 40$ &    8.36E-09(5.49E-09) & 1.00(1.07) &  $0.03^\ddag$$(0.05^\ddag)$ \\ 
 1800-2338$^a$ &   0.32 &  $1.10\pm0.10$ &    1 &    $ 40$ &    3.93E-10(2.16E-10) & 1.40(1.50) &  $0.71^\ddag$$(2.17)$ \\ 
 1809-2328 &   0.16 &  $1.06\pm0.08$ &    0 &    $ 40$ &    3.40E-10(2.10E-10) & 1.36(1.44) &  $0.88$$(2.29)$ \\ 
 1810-1032 &   0.39$^*$ &  $1.29\pm0.16$ &   12 &    $ 42$ &    6.17E-11(2.35E-11) & 1.59(1.75) &  $ 28$$(193)$ \\ 
 1812-1316 &   0.39 &  $1.29\pm0.11$ &   10 &    $ 41$ &    8.91E-11(4.59E-11) & 1.59(1.70) &  $ 13$$( 49)$ \\ 
 1824-1514 &   0.52 &  $1.19\pm0.18$ &    8 &    $ 41$ &    1.28E-10(4.34E-11) & 1.49(1.67) &  $6.28$$( 55)$ \\ 
 1826-1302 &   0.46 &  $1.00\pm0.11$ &   10 &    $ 41$ &    7.49E-10(3.86E-10) & 1.30(1.41) &  $0.37^\ddag$$(0.72^\ddag)$ \\ 
\multicolumn{7}{l}{{\it continued on the next page}}\\
\hline
\end{tabular}
}
\end{table}
\addtocounter{table}{-1}

\begin{table}[hbt]
\caption{ (part 2) Observability of unidentified and tentatively identified EGRET sources for project HESS I}
{\small
\begin{tabular}{l@{\hspace{0.15cm}}c@{\hspace{0.15cm}}c@{\hspace{0.15cm}}c@{\hspace{0.15cm}}c@{\hspace{0.15cm}}c@{\hspace{0.15cm}}c@{\hspace{0.15cm}}c}
\hline
\multicolumn{3}{c}{3EG} & \multicolumn{5}{|c}{ HESS I observability} \\
object  & \multicolumn{1}{c}{$\theta_{95}$$^b$}  
	& \multicolumn{1}{c}{$\alpha$$^c$}  
	& $\vartheta_{\mathrm{min}}$$^d$ 
	& \multicolumn{1}{c}{$E_{\mathrm{thr}}$$^e$} 
	& \multicolumn{1}{c}{$F(E_{\mathrm{thr}})$$^f$} 
	& \multicolumn{1}{c}{$\alpha(E_{\mathrm{thr}})$$^g$} 
	& \multicolumn{1}{c}{$T_{5\sigma}$$^h$}  \\ 
name    &\multicolumn{1}{c}{$[^\circ]$} 
	&  
	& \multicolumn{1}{c}{$[^\circ]$} 
	& \multicolumn{1}{c}{$[$GeV$]$} 
	& \multicolumn{1}{c}{$[$cm$^{-2}$s$^{-1}$$]$} 
	&  
	& \multicolumn{1}{c}{$[$h$]$}  \\
\hline
 1837-0606 &   0.19 &  $0.82\pm0.14$ &   17 &    $ 44$ &    1.54E-09(6.59E-10) & 1.12(1.26) &  $0.18^\ddag$$(0.42^\ddag)$ \\ 
 1850-2652 &   1.00 &  $1.29\pm0.45$ &    4 &    $ 40$ &    1.83E-10(1.23E-11) & 1.59(2.04) &  $3.03$$(>500)$ \\ 
 1856+0114$^a$ &   0.19 &  $0.93\pm0.10$ &   24 &    $ 48$ &    9.55E-10(5.15E-10) & 1.23(1.33) &  $0.29^\ddag$$(0.54^\ddag)$ \\ 
 1928+1733 &   0.75 &  $1.23\pm0.32$ &   41 &    $ 69$ &    1.99E-10(2.46E-11) & 1.53(1.85) &  $4.42$$(290)$ \\ 
 1958+2909 &   0.57 &  $0.85\pm0.20$ &   52 &    $106$ &    2.48E-10(6.17E-11) & 1.30(1.50) &  $4.36$$( 71)$ \\ 
\hline
\end{tabular}
}
\begin{tablenotes}
For definitions see the end of table \ref{tab-agenda-veritas}.
\end{tablenotes}
\end{table}

\begin{table}[hbt]
\caption{\label{tab-agenda-magic} Observability of unidentified and tentatively identified EGRET sources for project MAGIC I}
{\small
\begin{tabular}{l@{\hspace{0.15cm}}c@{\hspace{0.15cm}}c@{\hspace{0.15cm}}c@{\hspace{0.15cm}}c@{\hspace{0.15cm}}c@{\hspace{0.15cm}}c@{\hspace{0.15cm}}c}
\hline
\multicolumn{3}{c}{3EG} & \multicolumn{5}{|c}{ MAGIC I observability} \\
object  & \multicolumn{1}{c}{$\theta_{95}$$^b$}  
	& \multicolumn{1}{c}{$\alpha$$^c$}  
	& $\vartheta_{\mathrm{min}}$$^d$ 
	& \multicolumn{1}{c}{$E_{\mathrm{thr}}$$^e$} 
	& \multicolumn{1}{c}{$F(E_{\mathrm{thr}})$$^f$} 
	& \multicolumn{1}{c}{$\alpha(E_{\mathrm{thr}})$$^g$} 
	& \multicolumn{1}{c}{$T_{5\sigma}$$^h$}  \\ 
name    &\multicolumn{1}{c}{$[^\circ]$} 
	&  
	& \multicolumn{1}{c}{$[^\circ]$} 
	& \multicolumn{1}{c}{$[$GeV$]$} 
	& \multicolumn{1}{c}{$[$cm$^{-2}$s$^{-1}$$]$} 
	&  
	& \multicolumn{1}{c}{$[$h$]$}  \\
\hline
 0010+7309$^a$ &   0.24 &  $0.85\pm0.10$ &   44 &    $ 58$ &    7.87E-10(4.16E-10) & 1.15(1.25) &  $0.35^\ddag$$(0.67^\ddag)$ \\ 
 0215+1123 &   1.06 &  $1.03\pm0.62$ &   18 &    $ 33$ &    2.27E-10(6.22E-12) & 1.33(1.95) &  $1.23^\ddag$$(>500)$ \\ 
 0229+6151 &   0.49 &  $1.29\pm0.18$ &   33 &    $ 43$ &    7.07E-11(2.38E-11) & 1.59(1.77) &  $8.17$$( 72)$ \\ 
 0241+6103$^a$ &   0.18 &  $1.21\pm0.07$ &   32 &    $ 42$ &    2.15E-10(1.41E-10) & 1.51(1.58) &  $1.29^\ddag$$(2.01)$ \\ 
 0323+5122 &   0.55 &  $1.38\pm0.41$ &   22 &    $ 35$ &    3.67E-11(3.32E-12) & 1.68(2.09) &  $ 25$$(>500)$ \\ 
 0348+3510 &   0.74 &  $1.16\pm0.27$ &    6 &    $ 30$ &    7.70E-11(1.65E-11) & 1.46(1.73) &  $4.91$$(108)$ \\ 
 0426+1333 &   0.45$^*$ &  $1.17\pm0.25$ &   15 &    $ 32$ &    8.09E-11(1.91E-11) & 1.47(1.72) &  $4.74$$( 85)$ \\ 
 0439+1555 &   0.92 &  $1.27\pm0.44$ &   13 &    $ 32$ &    1.44E-10(1.14E-11) & 1.57(2.01) &  $1.93^\ddag$$(233)$ \\ 
 0510+5545 &   0.71 &  $1.19\pm0.20$ &   27 &    $ 38$ &    8.72E-11(2.66E-11) & 1.49(1.69) &  $4.75$$( 51)$ \\ 
 0613+4201 &   0.57 &  $0.92\pm0.26$ &   13 &    $ 32$ &    2.26E-10(5.07E-11) & 1.22(1.48) &  $1.23^\ddag$$( 12)$ \\ 
 0617+2238$^a$ &   0.13 &  $1.01\pm0.06$ &    6 &    $ 30$ &    8.11E-10(5.75E-10) & 1.31(1.37) &  $0.34^\ddag$$(0.48^\ddag)$ \\ 
 0628+1847 &   0.57 &  $1.30\pm0.10$ &   10 &    $ 31$ &    6.96E-11(3.92E-11) & 1.60(1.70) &  $6.13$$( 19)$ \\ 
 0631+0642 &   0.46 &  $1.06\pm0.15$ &   22 &    $ 35$ &    2.49E-10(1.03E-10) & 1.36(1.51) &  $1.12^\ddag$$(3.15)$ \\ 
 0634+0521 &   0.67 &  $1.03\pm0.26$ &   24 &    $ 36$ &    2.89E-10(6.26E-11) & 1.33(1.59) &  $0.96^\ddag$$(8.75)$ \\ 
 0808+4844$^a$ &   0.72 &  $1.15\pm0.45$ &   20 &    $ 34$ &    6.47E-11(4.71E-12) & 1.45(1.90) &  $7.76$$(>500)$ \\ 
 0812-0646$^a$ &   0.72 &  $1.34\pm0.29$ &   36 &    $ 46$ &    3.12E-11(5.29E-12) & 1.64(1.93) &  $ 45$$(>500)$ \\ 
 0910+6556 &   0.86 &  $1.20\pm0.26$ &   37 &    $ 47$ &    5.07E-11(1.02E-11) & 1.50(1.76) &  $ 18$$(430)$ \\ 
 0917+4427$^a$ &   0.56 &  $1.19\pm0.14$ &   15 &    $ 32$ &    7.10E-11(3.16E-11) & 1.49(1.63) &  $6.15$$( 31)$ \\ 
 1009+4855$^a$ &   0.75$^*$ &  $0.90\pm0.37$ &   20 &    $ 34$ &    1.48E-10(1.71E-11) & 1.20(1.57) &  $1.88^\ddag$$(112)$ \\ 
\multicolumn{7}{l}{{\it continued on the next page}}\\
\hline
\end{tabular}
}
\end{table}
\addtocounter{table}{-1}
\begin{table}[hbt]
\caption{ (part 2) Observability of unidentified and tentatively identified EGRET sources for project MAGIC I}
{\small
\begin{tabular}{l@{\hspace{0.15cm}}c@{\hspace{0.15cm}}c@{\hspace{0.15cm}}c@{\hspace{0.15cm}}c@{\hspace{0.15cm}}c@{\hspace{0.15cm}}c@{\hspace{0.15cm}}c}
\hline
\multicolumn{3}{c}{3EG} & \multicolumn{5}{|c}{ MAGIC I observability} \\
object  & \multicolumn{1}{c}{$\theta_{95}$$^b$}  
	& \multicolumn{1}{c}{$\alpha$$^c$}  
	& $\vartheta_{\mathrm{min}}$$^d$ 
	& \multicolumn{1}{c}{$E_{\mathrm{thr}}$$^e$} 
	& \multicolumn{1}{c}{$F(E_{\mathrm{thr}})$$^f$} 
	& \multicolumn{1}{c}{$\alpha(E_{\mathrm{thr}})$$^g$} 
	& \multicolumn{1}{c}{$T_{5\sigma}$$^h$}  \\ 
name    &\multicolumn{1}{c}{$[^\circ]$} 
	&  
	& \multicolumn{1}{c}{$[^\circ]$} 
	& \multicolumn{1}{c}{$[$GeV$]$} 
	& \multicolumn{1}{c}{$[$cm$^{-2}$s$^{-1}$$]$} 
	&  
	& \multicolumn{1}{c}{$[$h$]$}  \\
\hline
 1234-1318 &   0.76 &  $1.09\pm0.24$ &   42 &    $ 55$ &    9.48E-11(2.09E-11) & 1.39(1.63) &  $5.86$$(121)$ \\ 
 1323+2200$^a$ &   0.47 &  $0.86\pm0.35$ &    7 &    $ 30$ &    6.71E-10(9.07E-11) & 1.16(1.51) &  $0.41^\ddag$$(3.56)$ \\ 
 1337+5029 &   0.72 &  $0.83\pm0.29$ &   21 &    $ 35$ &    3.50E-10(6.42E-11) & 1.13(1.42) &  $0.79^\ddag$$(8.07)$ \\ 
 1631-1018 &   0.72 &  $1.20\pm0.27$ &   39 &    $ 50$ &    3.19E-11(5.96E-12) & 1.50(1.77) &  $ 47$$(>500)$ \\ 
 1719-0430 &   0.44 &  $1.20\pm0.24$ &   34 &    $ 43$ &    5.09E-11(1.19E-11) & 1.50(1.74) &  $ 16$$(294)$ \\ 
 1736-2908 &   0.62 &  $1.18\pm0.12$ &   58 &    $108$ &    4.67E-11(2.02E-11) & 1.63(1.75) &  $ 47$$(253)$ \\ 
 1744-3011 &   0.32 &  $1.17\pm0.08$ &   59 &    $114$ &    5.64E-11(3.21E-11) & 1.62(1.70) &  $ 35$$(107)$ \\ 
 1746-2851 &   0.13 &  $0.70\pm0.07$ &   58 &    $106$ &    3.16E-09(1.94E-09) & 1.15(1.22) &  $0.09^\ddag$$(0.14^\ddag)$ \\ 
 1800-2338$^a$ &   0.32 &  $1.10\pm0.10$ &   53 &    $ 82$ &    1.45E-10(7.42E-11) & 1.40(1.50) &  $3.71$$( 14)$ \\ 
 1809-2328 &   0.16 &  $1.06\pm0.08$ &   52 &    $ 81$ &    1.31E-10(7.64E-11) & 1.36(1.44) &  $4.55$$( 13)$ \\ 
 1810-1032 &   0.39$^*$ &  $1.29\pm0.16$ &   40 &    $ 50$ &    4.60E-11(1.70E-11) & 1.59(1.75) &  $ 23$$(167)$ \\ 
 1812-1316 &   0.39 &  $1.29\pm0.11$ &   42 &    $ 55$ &    5.66E-11(2.83E-11) & 1.59(1.70) &  $ 16$$( 66)$ \\ 
 1824+3441 &   0.82 &  $1.03\pm0.50$ &    6 &    $ 30$ &    4.05E-10(2.33E-11) & 1.33(1.83) &  $0.69^\ddag$$( 54)$ \\ 
 1824-1514 &   0.52 &  $1.19\pm0.18$ &   44 &    $ 58$ &    7.48E-11(2.38E-11) & 1.49(1.67) &  $ 10$$( 99)$ \\ 
 1826-1302 &   0.46 &  $1.00\pm0.11$ &   42 &    $ 54$ &    5.22E-10(2.61E-10) & 1.30(1.41) &  $0.53^\ddag$$(1.06^\ddag)$ \\ 
 1835+5918 &   0.15 &  $0.69\pm0.07$ &   30 &    $ 40$ &    4.50E-09(2.96E-09) & 0.99(1.06) &  $0.06^\ddag$$(0.09^\ddag)$ \\ 
 1837-0423 &   0.52 &  $1.71\pm0.44$ &   33 &    $ 43$ &    4.44E-11(3.08E-12) & 2.01(2.45) &  $ 21$$(>500)$ \\ 
 1837-0606 &   0.19 &  $0.82\pm0.14$ &   35 &    $ 45$ &    1.50E-09(6.38E-10) & 1.12(1.26) &  $0.19^\ddag$$(0.44^\ddag)$ \\ 
 1850-2652 &   1.00 &  $1.29\pm0.45$ &   56 &    $ 95$ &    4.64E-11(2.12E-12) & 1.59(2.04) &  $ 43$$(>500)$ \\ 
 1856+0114$^a$ &   0.19 &  $0.93\pm0.10$ &   28 &    $ 38$ &    1.26E-09(6.97E-10) & 1.23(1.33) &  $0.22^\ddag$$(0.40^\ddag)$ \\ 
 1903+0550$^a$ &   0.64 &  $1.38\pm0.17$ &   23 &    $ 35$ &    9.10E-11(3.35E-11) & 1.68(1.85) &  $4.12$$( 30)$ \\ 
 1928+1733 &   0.75 &  $1.23\pm0.32$ &   11 &    $ 31$ &    6.75E-10(1.07E-10) & 1.53(1.85) &  $0.41^\ddag$$(2.60)$ \\ 
 1958+2909 &   0.57 &  $0.85\pm0.20$ &    0 &    $ 30$ &    1.07E-09(3.43E-10) & 1.15(1.35) &  $0.26^\ddag$$(0.81^\ddag)$ \\ 
 2016+3657 &   0.55 &  $1.09\pm0.11$ &    8 &    $ 31$ &    3.43E-10(1.83E-10) & 1.39(1.50) &  $0.81^\ddag$$(1.52^\ddag)$ \\ 
 2020+4017$^a$ &   0.16 &  $1.08\pm0.04$ &   11 &    $ 31$ &    1.26E-09(1.00E-09) & 1.38(1.42) &  $0.22^\ddag$$(0.28^\ddag)$ \\ 
 2021+3716 &   0.30 &  $0.86\pm0.10$ &    8 &    $ 31$ &    2.18E-09(1.23E-09) & 1.16(1.26) &  $0.13^\ddag$$(0.23^\ddag)$ \\ 
 2022+4317 &   0.72 &  $1.31\pm0.19$ &   14 &    $ 32$ &    6.47E-11(2.16E-11) & 1.61(1.80) &  $7.34$$( 66)$ \\ 
 2027+3429 &   0.77 &  $1.28\pm0.15$ &    6 &    $ 30$ &    8.77E-11(3.72E-11) & 1.58(1.73) &  $3.78$$( 21)$ \\ 
 2033+4118 &   0.28 &  $0.96\pm0.10$ &   12 &    $ 31$ &    1.47E-09(8.26E-10) & 1.26(1.36) &  $0.19^\ddag$$(0.34^\ddag)$ \\ 
 2035+4441 &   0.54 &  $1.08\pm0.26$ &   16 &    $ 32$ &    2.83E-10(6.29E-11) & 1.38(1.64) &  $0.98^\ddag$$(7.85)$ \\ 
 2046+0933 &   0.60$^*$ &  $1.22\pm0.51$ &   19 &    $ 34$ &    8.42E-11(4.33E-12) & 1.52(2.03) &  $4.57$$(>500)$ \\ 
 2100+6012$^a$ &   0.48 &  $1.21\pm0.25$ &   31 &    $ 41$ &    6.33E-11(1.41E-11) & 1.51(1.76) &  $9.84$$(199)$ \\ 
 2206+6602$^a$ &   0.88 &  $1.29\pm0.26$ &   37 &    $ 47$ &    3.87E-11(7.80E-12) & 1.59(1.85) &  $ 30$$(>500)$ \\ 
 2227+6122 &   0.46 &  $1.24\pm0.14$ &   32 &    $ 42$ &    1.06E-10(4.55E-11) & 1.54(1.68) &  $3.59$$( 20)$ \\ 
 2248+1745 &   0.94 &  $1.11\pm0.39$ &   11 &    $ 31$ &    1.11E-10(1.18E-11) & 1.41(1.80) &  $2.51^\ddag$$(215)$ \\ 
 2255+1943$^a$ &   0.87$^*$ &  $1.36\pm0.61$ &    9 &    $ 31$ &    1.30E-10(3.93E-12) & 1.66(2.27) &  $2.14^\ddag$$(>500)$ \\ 
 2314+4426 &   0.78 &  $1.34\pm0.32$ &   15 &    $ 32$ &    8.74E-11(1.38E-11) & 1.64(1.96) &  $4.06$$(164)$ \\ 
 2352+3752$^a$ &   0.94 &  $1.47\pm0.68$ &    9 &    $ 31$ &    4.18E-11(8.50E-13) & 1.77(2.45) &  $ 17$$(>500)$ \\ 
\hline
\end{tabular}
}
\begin{tablenotes}
For definitions see the end of table \ref{tab-agenda-veritas}.
\end{tablenotes}
\end{table}

\begin{table}[hbt]
\caption{\label{tab-agenda-veritas} Observability of unidentified and tentatively identified EGRET sources for project VERITAS}
{\small
\begin{tabular}{l@{\hspace{0.15cm}}c@{\hspace{0.15cm}}c@{\hspace{0.15cm}}c@{\hspace{0.15cm}}c@{\hspace{0.15cm}}c@{\hspace{0.15cm}}c@{\hspace{0.15cm}}c}
\hline
\multicolumn{3}{c}{3EG} & \multicolumn{5}{|c}{ VERITAS observability} \\
object  & \multicolumn{1}{c}{$\theta_{95}$$^b$}  
	& \multicolumn{1}{c}{$\alpha$$^c$}  
	& $\vartheta_{\mathrm{min}}$$^d$ 
	& \multicolumn{1}{c}{$E_{\mathrm{thr}}$$^e$} 
	& \multicolumn{1}{c}{$F(E_{\mathrm{thr}})$$^f$} 
	& \multicolumn{1}{c}{$\alpha(E_{\mathrm{thr}})$$^g$} 
	& \multicolumn{1}{c}{$T_{5\sigma}$$^h$}  \\ 
name    &\multicolumn{1}{c}{$[^\circ]$} 
	&  
	& \multicolumn{1}{c}{$[^\circ]$} 
	& \multicolumn{1}{c}{$[$GeV$]$} 
	& \multicolumn{1}{c}{$[$cm$^{-2}$s$^{-1}$$]$} 
	&  
	& \multicolumn{1}{c}{$[$h$]$}  \\
\hline
 0010+7309$^a$ &   0.24 &  $0.85\pm0.10$ &   41 &    $141$ &    2.70E-10(1.31E-10) & 1.30(1.40) &  $1.03^\ddag$$(2.12^\ddag)$ \\ 
 0215+1123 &   1.06 &  $1.03\pm0.62$ &   21 &    $ 91$ &    5.86E-11(8.55E-13) & 1.33(1.95) &  $4.74^\ddag$$(>500)$ \\ 
 0229+6151 &   0.49 &  $1.29\pm0.18$ &   30 &    $106$ &    1.63E-11(4.65E-12) & 1.74(1.92) &  $ 25$$(308)$ \\ 
 0241+6103$^a$ &   0.18 &  $1.21\pm0.07$ &   29 &    $105$ &    5.34E-11(3.28E-11) & 1.66(1.73) &  $5.20^\ddag$$(8.47^\ddag)$ \\ 
 0348+3510 &   0.74 &  $1.16\pm0.27$ &    3 &    $ 80$ &    1.86E-11(3.06E-12) & 1.46(1.73) &  $ 15^\ddag$$(>500)$ \\ 
 0426+1333 &   0.45$^*$ &  $1.17\pm0.25$ &   18 &    $ 89$ &    1.83E-11(3.34E-12) & 1.47(1.72) &  $ 17$$(497)$ \\ 
 0439+1555 &   0.92 &  $1.27\pm0.44$ &   16 &    $ 87$ &    2.95E-11(1.51E-12) & 1.57(2.01) &  $9.41^\ddag$$(>500)$ \\ 
 0510+5545 &   0.71 &  $1.19\pm0.20$ &   24 &    $ 96$ &    2.18E-11(5.52E-12) & 1.49(1.69) &  $ 13^\ddag$$(196)$ \\ 
 0613+4201 &   0.57 &  $0.92\pm0.26$ &   10 &    $ 82$ &    7.02E-11(1.22E-11) & 1.22(1.48) &  $3.96^\ddag$$( 34)$ \\ 
 0617+2238$^a$ &   0.13 &  $1.01\pm0.06$ &    9 &    $ 82$ &    2.20E-10(1.47E-10) & 1.31(1.37) &  $1.26^\ddag$$(1.89^\ddag)$ \\ 
 0628+1847 &   0.57 &  $1.30\pm0.10$ &   13 &    $ 84$ &    1.40E-11(7.14E-12) & 1.60(1.70) &  $ 27$$(104)$ \\ 
 0631+0642 &   0.46 &  $1.06\pm0.15$ &   25 &    $ 98$ &    6.16E-11(2.19E-11) & 1.36(1.51) &  $4.51^\ddag$$( 13)$ \\ 
 0634+0521 &   0.67 &  $1.03\pm0.26$ &   27 &    $100$ &    7.34E-11(1.22E-11) & 1.48(1.74) &  $3.78^\ddag$$( 42)$ \\ 
 0808+4844$^a$ &   0.72 &  $1.15\pm0.45$ &   17 &    $ 87$ &    1.64E-11(7.80E-13) & 1.45(1.90) &  $ 20$$(>500)$ \\ 
 0910+6556 &   0.86 &  $1.20\pm0.26$ &   34 &    $116$ &    1.27E-11(2.03E-12) & 1.65(1.91) &  $ 45$$(>500)$ \\ 
 0917+4427$^a$ &   0.56 &  $1.19\pm0.14$ &   12 &    $ 84$ &    1.71E-11(6.67E-12) & 1.49(1.63) &  $ 18$$(118)$ \\ 
 1009+4855$^a$ &   0.75$^*$ &  $0.90\pm0.37$ &   17 &    $ 87$ &    4.74E-11(3.87E-12) & 1.20(1.57) &  $5.86^\ddag$$(365)$ \\ 
 1234-1318 &   0.76 &  $1.09\pm0.24$ &   45 &    $162$ &    1.96E-11(3.33E-12) & 1.54(1.78) &  $ 26$$(>500)$ \\ 
 1323+2200$^a$ &   0.47 &  $0.86\pm0.35$ &   10 &    $ 82$ &    2.11E-10(2.01E-11) & 1.16(1.51) &  $1.32^\ddag$ $( 14^\ddag)$ \\ 
 1337+5029 &   0.72 &  $0.83\pm0.29$ &   18 &    $ 89$ &    1.21E-10(1.68E-11) & 1.13(1.42) &  $2.30^\ddag$$( 20)$ \\ 
 1800-2338$^a$ &   0.32 &  $1.10\pm0.10$ &   56 &    $251$ &    2.61E-11(1.19E-11) & 1.55(1.65) &  $ 23$$(110)$ \\ 
 1809-2328 &   0.16 &  $1.06\pm0.08$ &   55 &    $249$ &    2.47E-11(1.32E-11) & 1.51(1.59) &  $ 26$$( 89)$ \\ 
 1824+3441 &   0.82 &  $1.03\pm0.50$ &    3 &    $ 80$ &    1.11E-10(3.92E-12) & 1.33(1.83) &  $2.50^\ddag$$(326)$ \\ 
 1826-1302 &   0.46 &  $1.00\pm0.11$ &   45 &    $160$ &    1.19E-10(5.31E-11) & 1.45(1.56) &  $2.33^\ddag$$(5.24^\ddag)$ \\ 
 1835+5918 &   0.15 &  $0.69\pm0.07$ &   27 &    $101$ &    1.80E-09(1.11E-09) & 1.14(1.21) &  $0.15^\ddag$$(0.25^\ddag)$ \\ 
 1837-0606 &   0.19 &  $0.82\pm0.14$ &   38 &    $129$ &    4.41E-10(1.62E-10) & 1.27(1.41) &  $0.63^\ddag$$(1.72^\ddag)$ \\ 
 1856+0114$^a$ &   0.19 &  $0.93\pm0.10$ &   31 &    $108$ &    3.48E-10(1.73E-10) & 1.38(1.48) &  $0.80^\ddag$$(1.61^\ddag)$ \\ 
 1903+0550$^a$ &   0.64 &  $1.38\pm0.17$ &   26 &    $ 99$ &    1.62E-11(5.00E-12) & 1.68(1.85) &  $ 24$$(249)$ \\ 
 1928+1733 &   0.75 &  $1.23\pm0.32$ &   14 &    $ 85$ &    1.45E-10(1.67E-11) & 1.53(1.85) &  $1.92^\ddag$$( 19)$ \\ 
 1958+2909 &   0.57 &  $0.85\pm0.20$ &    3 &    $ 80$ &    3.47E-10(9.10E-11) & 1.15(1.35) &  $0.80^\ddag$$(3.05^\ddag)$ \\ 
 2016+3657 &   0.55 &  $1.09\pm0.11$ &    5 &    $ 81$ &    8.92E-11(4.27E-11) & 1.39(1.50) &  $3.11^\ddag$$(6.50^\ddag)$ \\ 
 2020+4017$^a$ &   0.16 &  $1.08\pm0.04$ &    8 &    $ 82$ &    3.34E-10(2.55E-10) & 1.38(1.42) &  $0.83^\ddag$$(1.09^\ddag)$ \\ 
 2021+3716 &   0.30 &  $0.86\pm0.10$ &    5 &    $ 81$ &    7.08E-10(3.62E-10) & 1.16(1.26) &  $0.39^\ddag$$(0.77^\ddag)$ \\ 
 2022+4317 &   0.72 &  $1.31\pm0.19$ &   11 &    $ 83$ &    1.38E-11(3.86E-12) & 1.61(1.80) &  $ 27$$(349)$ \\ 
 2027+3429 &   0.77 &  $1.28\pm0.15$ &    2 &    $ 80$ &    1.88E-11(6.91E-12) & 1.58(1.73) &  $ 15^\ddag$$(105)$ \\ 
 2033+4118 &   0.28 &  $0.96\pm0.10$ &    9 &    $ 82$ &    4.37E-10(2.24E-10) & 1.26(1.36) &  $0.64^\ddag$$(1.24^\ddag)$ \\ 
 2035+4441 &   0.54 &  $1.08\pm0.26$ &   13 &    $ 84$ &    7.58E-11(1.32E-11) & 1.38(1.64) &  $3.67^\ddag$$( 30)$ \\ 
 2046+0933 &   0.60$^*$ &  $1.22\pm0.51$ &   22 &    $ 94$ &    1.78E-11(5.45E-13) & 1.52(2.03) &  $ 18$$(>500)$ \\ 
\multicolumn{7}{l}{{\it continued on the next page}}\\
\hline
\end{tabular}
}
\end{table}
\addtocounter{table}{-1}

\begin{table}[hbt]
\caption{(part 2) Observability of unidentified and tentatively identified EGRET sources for project VERITAS}
{\small
\begin{tabular}{l@{\hspace{0.15cm}}c@{\hspace{0.15cm}}c@{\hspace{0.15cm}}c@{\hspace{0.15cm}}c@{\hspace{0.15cm}}c@{\hspace{0.15cm}}c@{\hspace{0.15cm}}c}
\hline
\multicolumn{3}{c}{3EG} & \multicolumn{5}{|c}{ VERITAS observability} \\
object  & \multicolumn{1}{c}{$\theta_{95}$$^b$}  
	& \multicolumn{1}{c}{$\alpha$$^c$}  
	& $\vartheta_{\mathrm{min}}$$^d$ 
	& \multicolumn{1}{c}{$E_{\mathrm{thr}}$$^e$} 
	& \multicolumn{1}{c}{$F(E_{\mathrm{thr}})$$^f$} 
	& \multicolumn{1}{c}{$\alpha(E_{\mathrm{thr}})$$^g$} 
	& \multicolumn{1}{c}{$T_{5\sigma}$$^h$}  \\ 
name    &\multicolumn{1}{c}{$[^\circ]$} 
	&  
	& \multicolumn{1}{c}{$[^\circ]$} 
	& \multicolumn{1}{c}{$[$GeV$]$} 
	& \multicolumn{1}{c}{$[$cm$^{-2}$s$^{-1}$$]$} 
	&  
	& \multicolumn{1}{c}{$[$h$]$}  \\
\hline
 2100+6012$^a$ &   0.48 &  $1.21\pm0.25$ &   28 &    $103$ &    1.57E-11(2.77E-12) & 1.66(1.91) &  $ 26$$(>500)$ \\ 
 2227+6122 &   0.46 &  $1.24\pm0.14$ &   29 &    $105$ &    2.56E-11(9.65E-12) & 1.69(1.83) &  $ 11^\ddag$$( 71)$ \\ 
 2248+1745 &   0.94 &  $1.11\pm0.39$ &   14 &    $ 85$ &    2.69E-11(1.93E-12) & 1.41(1.80) &  $ 10^\ddag$$(>500)$ \\ 
 2255+1943$^a$ &   0.87$^*$ &  $1.36\pm0.61$ &   12 &    $ 84$ &    2.46E-11(4.06E-13) & 1.66(2.27) &  $ 11^\ddag$$(>500)$ \\ 
 2314+4426 &   0.78 &  $1.34\pm0.32$ &   12 &    $ 84$ &    1.83E-11(2.12E-12) & 1.64(1.96) &  $ 16$$(>500)$ \\ 
\hline
\end{tabular}
}
\begin{tablenotes}
$^a$ The object has already been tentatively identified in 3EG (Hartman et al. \protect\cite{egret3rdcat}).

$^b$ The radius of the 95\% confidence contour in the position probability map in 3EG.

$^c$ The integral spectral index at 0.1 GeV (see equ. \protect\ref{equ-extraflux}) with statistical error from 3EG.

$^d$ The minimum zenith angle, see equ. \protect\ref{equ-thetamin}.

$^e$ The IACTs minimum energy threshold for this source, see equ. \protect\ref{equ-ethr}.

$^f$ The expected flux at the minimum energy threshold, see equ. \protect\ref{equ-extraflux}; 
     in brackets the flux which is obtained from an extrapolation using an initial spectrum at 0.1 GeV
     steeper by one standard deviation.

$^g$ The integral spectral index at the IACTs minimum energy threshold for this source, see equ. \protect\ref{equ-extraflux};
   in brackets the index obtained if the spectrum at 0.1 GeV was steeper by one standard deviation.

$^h$ The observation time to obstain a detection with 5 $\sigma$ significance; in brackets the observation time
  necessary if the spectrum at 0.1 GeV was steeper by one standard deviation.

$^*$ The shape of the position probability map is irregular (see 3EG).

$^\ddag$ The detection is photon flux limited: the observation time was increased such that 100 gammas are detected.
\end{tablenotes} 

\end{table}

\begin{table}[hbt]
\caption{\label{tab-agenda} The prime candidates for EGRET-UNID observations by the next-generation Imaging Cherenkov Telescopes}
{\small
\begin{tabular*}{\textwidth}{@{\extracolsep{\fill}}llcccc}
\hline
3EG name & tentative ID & $\vartheta_{\mathrm{min}}$(C)$^a$ & $\vartheta_{\mathrm{min}}$(H)$^b$ & $\vartheta_{\mathrm{min}}$(M)$^c$ & $\vartheta_{\mathrm{min}}$(V)$^d$ \\
 & & $[^\circ]$ & $[^\circ]$ & $[^\circ]$ & $[^\circ]$ \\
\hline
 0010+7309 & SNR CTA\,1    &  -   &  -   & 44    & 41    \\ 
 0241+6103 & LS\,I\,+61$^\circ$303    &  -   &  -   & 32    & 29    \\ 
 0613+4201 &   -      &  -   &  -   & 13    & 10    \\ 
 0617+2238 & SNR IC443   &  -   & 46    &  6    &  9    \\ 
 0628+1847 &   -      &  -   &  -   & 10    &  -   \\ 
 0631+0642 &   -      &  -   & 30    & 22    & 25    \\ 
 0634+0521 &   -      &  -   &  -   & 24    & 27    \\ 
 0827-4247 & SNR Pup A & 12    & 20    &  -   &  -   \\ 
 0841-4356 &   -      & 13    & 21    &  -   &  -   \\ 
 0848-4429 &   -      & 14    & 22    &  -   &  -   \\ 
 0917+4427 & QSO 0917+449    &  -   &  -   & 15    &  -   \\ 
 1027-5817 &   -      & 27    & 35    &  -   &  -   \\ 
 1048-5840 &   -      & 28    & 36    &  -   &  -   \\ 
 1323+2200 & EQ 1324+224   &  -   &  -   &  7    & 10    \\ 
 1337+5029 &   -      &  -   &  -   & 21    & 18    \\ 
 1410-6147 & SNR 312.4-00.4   & 31    & 39    &  -   &  -   \\ 
 1420-6038 &   -      & 30    & 38    &  -   &  -   \\ 
 1704-4732 &   -      & 17    & 25    &  -   &  -   \\ 
 1736-2908 &   -      &  2    &  6    &  -   &  -   \\ 
 1744-3011 &   -      &  1    &  7    &  -   &  -   \\ 
 1746-2851 &   -      &  2    &  6    & 58    &  -   \\ 
 1800-2338 & SNR W\,28      &  7    &  1    & 53    &  -   \\ 
 1809-2328 &   -      &  8    &  0    & 52    &  -   \\ 
 1812-1316 &   -      &  -   & 10    &  -   &  -   \\ 
 1826-1302 &   -      & 18    & 10    & 42    & 45    \\ 
 1835+5918 &   -      &  -   &  -   & 30    & 27    \\ 
 1837-0606 &   -      & 25    & 17    & 35    & 38    \\ 
 1856+0114 & SNR W\,44    & 32    & 24    & 28    & 31    \\ 
 1903+0550 & SNR 040.5-00.5    &  -   &  -   & 23    &  -   \\ 
 1928+1733 &   -      &  -   &  -   & 11    & 14    \\ 
 1958+2909 &   -      &  -   &  -   &  0    &  3    \\ 
 2016+3657 &   -      &  -   &  -   &  8    &  5    \\ 
 2020+4017 & SNR $\gamma$Cyg    &  -   &  -   & 11    &  8    \\ 
 2021+3716 &   -      &  -   &  -   &  8    &  5    \\ 
 2027+3429 &   -      &  -   &  -   &  6    &  -   \\ 
 2033+4118 &   -      &  -   &  -   & 12    &  9    \\ 
 2035+4441 &   -      &  -   &  -   & 16    & 13    \\ 
 2227+6122 &   -      &  -   &  -   & 32    &  -   \\ 
\hline
\end{tabular*}
}
\begin{tablenotes}
$^a$ the minimum zenith angle for observations with CANGAROO III

$^b$ the minimum zenith angle for observations with HESS I

$^c$ the minimum zenith angle for observations with MAGIC I

$^d$ the minimum zenith angle for observations with VERITAS
\end{tablenotes}
\end{table}

\end{document}